\documentclass[atoms,article,submit,pdftex,moreauthors]{Definitions/mdpipp}

\firstpage{1}
\pubvolume{1}
\issuenum{1}
\articlenumber{0}
\pubyear{2023}
\copyrightyear{2023}
\externaleditor{Academic Editor: Mark Edwards}
\datereceived{23 December 2022}
\daterevised{6 February 2023}
\dateaccepted{8 February 2023}
\datepublished{10 February 2023}
\hreflink{https://doi.org/} 
\pdfoutput=1

\usepackage[utf8]{inputenc}
\usepackage{amsmath}
\usepackage{braket}
\usepackage{graphicx}
\usepackage{color}

\newcommand{\Op}[1]{\ensuremath{\mathsf{\hat{#1}}}}

\newcommand{\Abs}[1]{\left|#1\right|}
\newcommand{\Norm}[1]{\left\lVert#1\right\rVert}

\newcommand{\ketbra}[2]{\ket{#1}\!\bra{#2}}

\newcommand{\target}{\text{tgt}}
\newcommand{\TLS}{TLS}  

\newcommand{\Rabi}{\text{Rabi}}
\newcommand{\pop}{\text{pop}}
\newcommand{\RAP}{\text{RAP}}

\newcommand{\microsec}{\textmu{}s}

\Title{Robust Optimized Pulse Schemes \\for Atomic Fountain Interferometry}
\TitleCitation{Robust Optimized Pulse Schemes for Atomic Fountain Interferometry}

\Author{Michael H. Goerz$^1$, Mark A. Kasevich$^2$, Vladimir S. Malinovsky$^{1,*}$}
\AuthorNames{Michael H. Goerz, Mark A. Kasevich and Vladimir S. Malinovsky}
\AuthorCitation{Goerz, M.H.; Kasevich, M.A.; Malinovsky, V.S.}

\address{%
$^1$ \quad DEVCOM Army Research Laboratory, Adelphi, MD 20783, USA.\\
$^2$ \quad Department of Physics, Stanford University, Stanford, CA 94305, USA.
}

\corres{Correspondence: vladimir.s.malinovsky.civ@army.mil}

\abstract{%
The robustness of an atomic fountain interferometer with respect to variations in the initial velocity of the atoms and deviations from the optimal pulse amplitude is examined. We numerically simulate the dynamics of an interferometer in momentum space with a maximum separation of $20\,\hbar k$ and map out the expected signal contrast depending on the variance of the initial velocity distribution and the value of the laser field amplitude. We show that an excitation scheme based on rapid adiabatic passage significantly enhances the expected signal contrast, compared to the commonly used scheme consisting of a series of $\pi/2$ and $\pi$ pulses. We demonstrate further substantial increase of the robustness by using optimal control theory to identify splitting and swapping pulses that perform well on an ensemble average of pulse amplitudes and velocities. Our results demonstrate the ability of optimal control to significantly enhance future implementations of atomic fountain interferometry.
}

\begin{document}

\section{Introduction}%
\label{sec:intro}

Atom interferometry~\cite{Berman1997,BaudonJPB1999,CroninRMP2009} is a core technology for quantum metrology and quantum sensing~\cite{DegenRMP2017}. Its operating principle exploits the wave nature of matter in the same way that classical optical interferometers exploit the wave nature of light. In the type of atom interferometer considered here, an initial wavepacket is split into two pathways that are separated spatially. These pathways are then mirrored and recombined. The recombination maps an accumulated phase difference between the two pathways into population at the output ports as an interference pattern.

Since atoms have mass, this relative phase is affected by gravitational fields, acceleration, and rotation. This enables not only applications in fundamental physics~\cite{DimopoulosPRD2008, DimopoulosPRD2008b, SchlippertPRL2014}, but also direct practical implementations of gradiometers, gyroscopes, and inertial navigation sensors~\cite{AbeQST2021,NarducciAPX2022}.
A second advantage, as well as a challenge of atomic matter waves, is that their wavelength is many orders of magnitude smaller than that of light, thus promising extremely high sensitivity, but making it difficult to implement ``beamsplitters'' and ``mirrors'' that can diffract at that scale. The highest precision can be reached through light-matter interaction~\cite{Shore2011,BermanMalinovsky2011}; specifically, the interaction of the atoms with counter-propagating laser fields.
There are different regimes in which the atoms can absorb or emit photons, imparting momentum kicks~\cite{HartmannPRA2020}. Most commonly, in the Raman regime, the change in momentum is associated with a change in the internal state of the atom~\cite{BordePLA1989,KasevichPRL1991,KasevichAPB1992,McGuirkPRL2000}, whereas in the Bragg regime, the atom remains in its internal ground state~\cite{ChebotayevJOSAB1985,GiltnerPRL1995,RaselPRL1995,DelhuilleAPB2002,MullerPRL2008}.

Here, we consider the setup of an atomic fountain interferometer (AFI)~\cite{MarchiM1982, KasevichON1989, KasevichPRL1989,MullerPRL2008}. We focus on the Bragg regime, which allows for higher momentum space separation and is less susceptible to ac-Stark and Zeeman shifts~\cite{AltinNJP2013}. A cloud of atoms with a well-defined initial momentum is launched into a 10~m tower~\cite{KovachyN2015}. After the launch, counter-propagating laser beams with appropriately controlled amplitude and phase impart momentum kicks onto the atoms to implement a momentum space beamsplitter, mirror, and recombination. The scale of the atomic fountain architecture allows for time of flight on the order of seconds and, thus, achieves the largest interferometric space-time area to date.

The contrast of the interferometer, particularly in the Bragg regime, is constrained by the spread in the initial momentum of the atoms~\cite{McGuirkPRL2000,SzigetiNJP2012} as well as by deviations in the laser field intensity~\cite{LuoEPJD2016}. Here, we quantify the expected contrast for different widths in the initial velocity distribution, and for variations in the pulse amplitude by $\pm 10\%$, for a sequence of $\pi/2$ and $\pi$ pulses.  We further analyze the robustness of analytical pulse schemes using rapid adiabatic passage (RAP)~\cite{PeikPRA1997,MalinovskyPRA2003}. These use linearly chirped laser pulses to transfer population between the momentum states.  In the context of AFI, they have been demonstrated to achieve large momentum space separation~\cite{MalinovskyPRA2003,KovachyPRA2012}. RAP is inherently robust to deviations in the pulse amplitude. We show here that the combination of RAP and $\pi/2$ and $\pi$ pulses significantly enhances the robustness of the full interferometric scheme.

Going beyond standard analytic pulse shapes, such as Gaussians, more elaborate shapes have been demonstrated to improve robustness~\cite{LuoEPJD2016, SaywellPRA2018}. Finding appropriate pulse shapes is the focus of optimal control theory (OCT)~\cite{BrumerShapiro2003,BrifNJP2010,Sola_AAMO2018}.  For atom interferometers in the Raman regime and two-level dynamics, the ability of optimal control to achieve robust pulses has been demonstrated~\cite{SaywellPRA2018,SaywellJPB2020,SaywellPRA2020}. Optimal control has also been applied to Bose-Einstein condensates in atom interferometers~\cite{vanFrankNC2014}. Here, we use Krotov's method of optimal control~\cite{TannorBookChapter1992,SomloiCP1993,ReichJCP2012,GoerzSPP2019} for an ensemble optimization~\cite{LiITAC2009,ChenPRA2014,GoerzPRA2014} to identify robust pulses for large momentum state separation in the Bragg regime. The ensemble optimization targets the average of a statistical ensemble of atoms with different initial velocities and experiencing different pulse amplitudes. We find that the most robust scheme overall uses optimized pulses driving transitions between the ground momentum state $\ket{0}$ and the first excited momentum state $\ket{1}$ corresponding to the momentum $2\hbar k$. These optimized pulses are combined with analytical RAP pulses to amplify the momentum state separation, to the momentum state $\ket{10}$ corresponding to $20 \hbar k$, in our example.

This paper is organized as follows. In Section~\ref{sec:model}, we describe the interaction of an atom with two counter-propagating laser beams inside an atomic fountain interferometer and derive its momentum space Hamiltonian in the Bragg regime. Section~\ref{sec:analytical} reviews two analytic pulse schemes: first, a train of $\pi/2$ and $\pi$ pulses, and second, a combination of $\pi/2$ and $\pi$ pulses and rapid adiabatic passage. Section~\ref{sec:robustness} analyzes the robustness of these schemes in terms of the expected contrast for varying pulse amplitudes and for initial velocities of varying uncertainties. In Section~\ref{sec:optrobust}, we describe the ensemble optimization approach used to improve the robustness of the analytical schemes. We analyze the resulting control fields and dynamics and compare the resulting robustness throughout the parameter landscape. Section~\ref{sec:conclusion} concludes.

\section{Model}%
\label{sec:model}

\begin{figure}[tb]
  \centering
  \includegraphics{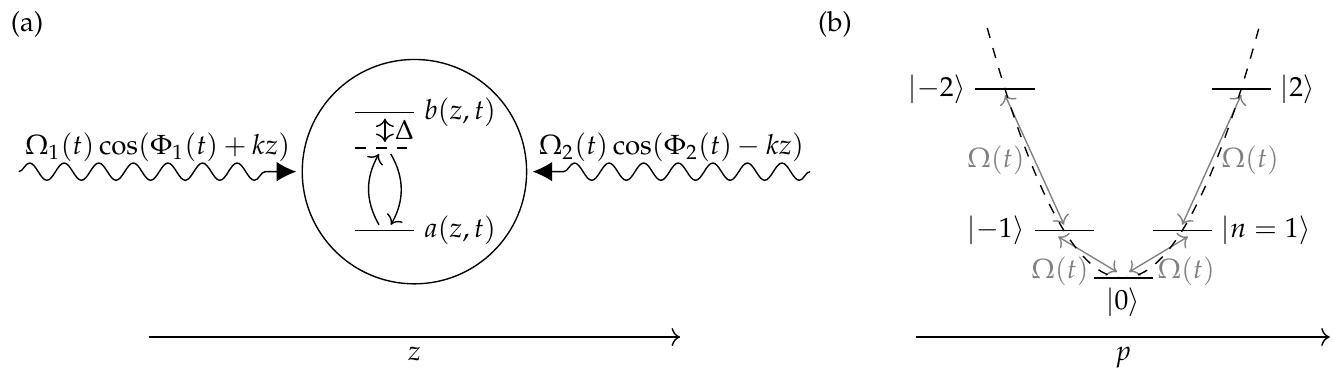}
  \caption{%
    \label{fig:system}
    Interaction of an atom with interferometer laser fields in the Bragg regime. (\textbf{a}) Two-level atom described as $\ket{\Psi(z, t)} = a(z,t) \ket{g} + b(z,t) \ket{e}$ in the coordinate representation with the two-level transition $\ket{g} \leftrightarrow \ket{e}$ off-resonantly driven by two fields counter-propagating along the $z$ axis, with a detuning $\Delta$ and time-dependent amplitudes $\Omega_1(t)$ and $\Omega_2(t)$ and time-dependent phases $\Phi_1(t)$ and $\Phi_2(t)$.
    (\textbf{b}) Momentum space ladder $\ket{n}$ corresponding to momentum $n \cdot 2 \hbar k$, with transitions between neighboring levels being driven by the effective pulse amplitude $\Omega(t)$.
  }
\end{figure}
In this paper, we consider an atomic fountain interferometer~\cite{KasevichPRL1989, MullerPRL2008,ChiowPRL2011} with light pulses in the Bragg regime~\cite{ChebotayevJOSAB1985,GiltnerPRL1995, RaselPRL1995}.
A cloud of ultracold Rubidium atoms with atomic mass $m$ is launched along the $z$ axis of a ten-meter tower~\cite{KovachyN2015}.
After the launch, the atoms are subjected to two counter-propagating laser fields with wave number $k = 2\pi/\lambda$ for a given laser wavelength $\lambda$.
These laser fields have tunable amplitude $\Omega_1(t)$, $\Omega_2(t)$ and frequency $\Phi_1(t)$, $\Phi_2(t)$ and are used to implement momentum space ``beamsplitters'' and ``mirrors''.
A schematic of the interaction for a single atom is shown in Figure~\ref{fig:system}~(a).

The laser fields off-resonantly drive an internal electronic transition, e.g., the Rubidium D2 line $5^2S_{1/2} \rightarrow 5^2P_{3/2}$. The atom is described by the wave function $\ket{\Psi(z,t)} = a(z, t) \ket{g} + b(z, t) \ket{e}$, where $\ket{g}$ and $\ket{e}$ are the $5^2S_{1/2}$ and $5^2P_{3/2}$ levels, respectively.
For a large detuning $\Delta$, the amplitude $b(z,t)$ of the excited state can be adiabatically eliminated.
This results in an effective two-photon field with amplitude and phase
\begin{equation}
  \Omega(t) = \frac{\Omega_1(t) \Omega_2(t)}{4 \Delta}\,, \qquad
  \varphi(t) = \Phi_1(t) - \Phi_2(t)\,,
\end{equation}
acting only on the ground state amplitude $a(z,t)$~\cite{GoerzSPIEO2021}.
A transformation from coordinate space to momentum space shows that the effective field can change the momentum only in discrete units of $2\hbar k$ relative to the rest frame defined for the initial momentum $p_0$~\cite{WichtPRA2005,KovachyPRA2010}.

We thus consider the entire dynamics of the atom in momentum space, where the ground state $\ket{0}$ corresponds to the rest frame momentum $p_0$, and the Hilbert space is defined in terms of levels $\ket{n}$ corresponding to the momentum $p_0 + n \cdot 2\hbar k$.
The Hamiltonian takes the form
\begin{equation}
  \label{eq:hamiltonian}
  \Op{H} =
    \left(\,\sum_{n=-\infty}^{\infty} E_n(t) \ketbra{n}{n} \right)
    - \left(\,
        \sum_{n=-\infty}^{\infty}\left(
         \mu\,\hbar\Omega(t) \ketbra{n}{n+1} + \mu\,\hbar\Omega^*(t) \ketbra{n+1}{n}
        \right)
      \right)\,.
\end{equation}
The time-dependent energy levels are
\begin{equation}
  \label{eq:E_n_exact}
  E_n(t)
  = n^2 \, \hbar\omega_k + \beta^2 \, \hbar\omega_k - \hbar \Omega_\Delta(t) + 2n \,\beta \, \hbar \omega_k + n \hbar \dot\varphi(t)
\end{equation}
with the two-photon recoil frequency $\omega_k = 2\hbar k^2/m$, the dimensionless initial momentum $\beta = p_0/2\hbar k$, and
the common light shift $\Omega_\Delta(t) = (\Omega_1^2(t) + \Omega_2^2(t))/4\Delta$.
We consider here $\beta < 1$ and neglect the term proportional to $\beta^2$. Moreover, as it does not depend on $n$, the common light shift $\Omega_\Delta(t)$ only contributes a global phase to the dynamics of the interferometer and can be dropped. Thus, we use
\begin{equation}
  \label{eq:E_n}
  E_n(t)
  \approx n^2 \, \hbar\omega_k + 2n \,\beta \, \hbar \omega_k + n \hbar \dot\varphi(t)
\end{equation}
in the remainder of the paper.

The energy levels $E_n(t)$ form the parabolic momentum ladder shown in Figure~\ref{fig:system}~(b). These levels are shifted by the angular frequency $\dot\varphi(t)$ of the driving field. An appropriate frequency can shift neighboring levels into resonance, which then allows the effective amplitude $\Omega(t)$ to transfer population. A nonzero value of $\beta$, that is, a nonzero momentum of the atom relative to the rest frame, would result in an effective detuning of this momentum space transition.

The factor $\mu$ in Equation~\eqref{eq:hamiltonian} accounts for deviations of the effective pulse amplitude $\Omega(t)$ from the optimal value, e.g., due to misalignment of the position of the atomic cloud relative to the cross section of the laser pulse. By default, we take $\mu = 1$. The effective amplitude $\Omega(t)$ is allowed to be complex-valued only in the context of optimal control. Optimizing the real and imaginary part of a complex-valued $\Omega(t)$ independently is equivalent to optimizing a real-valued field amplitude and phase. That is, the complex phase of $\Omega(t)$ expresses a deviation from the phase $\varphi(t)$ in Equation~\eqref{eq:E_n}.

For complete generality, we express the energies $E_n(t)$ and $\hbar\Omega(t)$ in units of $\hbar\omega_k$. Using the corresponding time unit of $1/\omega_k$ makes the Schrödinger equation dimensionless. For Rb-87, with a laser wavelength of $\lambda = 780$~nm, we have $\omega_k = 2 \pi \cdot 15.1$~kHz, and a corresponding time unit of approximately $10^{-5}$~s.

\section{Analytical Pulse Schemes}%
\label{sec:analytical}

A full atomic fountain interferometer ideally consists of the following steps:

\begin{enumerate}

  \item Split the initial state $\ket{0}$ into a superposition of $\ket{0}$ and $\ket{N}$. In the example here, we use $N=10$, corresponding to a momentum state separation of $20 \, \hbar k$.  This is the analog of a beamsplitter in a classic interferometer. The splitting may further be subdivided as:
    \begin{enumerate}[leftmargin=2.1em,labelsep=5mm]
      \item Perform an \emph{initial splitting}, from $\ket{0}$ to a superposition of $\ket{0}$ and $\ket{1}$.
      \item \emph{Amplify} the momentum state separation by transferring population from $\ket{1}$ to $\ket{N}$, resulting in a superposition of $\ket{0}$ and $\ket{N}$.
    \end{enumerate}

  \item Let the atoms evolve freely as they travel up the tower. During this time, an external gravitational field or acceleration may introduce a differential phase between the states $\ket{0}$ and $\ket{N}$.

  \item Swap the complex amplitudes of the $\ket{0}$ and $\ket{N}$ states. This is the analog of a mirror in a classic interferometer. It may be subdivided as in step 1:
    \begin{enumerate}[leftmargin=2.1em,labelsep=5mm]
      \item \emph{De-amplify} the population from $\ket{N}$ to $\ket{1}$, so that the interferometer is in a superposition of $\ket{0}$ and $\ket{1}$.
      \item \emph{Swap} the amplitude of $\ket{0}$ and $\ket{1}$.
      \item \emph{Amplify} the population from $\ket{1}$ to $\ket{N}$, which again brings the interferometer into a superposition of $\ket{0}$ and $\ket{N}$, with swapped amplitudes relative to the end of step 2.
    \end{enumerate}

  \item Let the atoms continue to evolve freely as they descend the tower, potentially accumulating a further differential phase.

  \item Recombine the state into a superposition of $\ket{0}$ and $\ket{1}$. This would be subdivided as:
    \begin{enumerate}[leftmargin=2.1em,labelsep=5mm]
      \item \emph{De-amplify} the population from $\ket{N}$ to $\ket{1}$, resulting in a superposition of $\ket{0}$ and $\ket{1}$.
      \item Coherently \emph{recombine} the amplitudes of $\ket{0}$ and $\ket{1}$ by applying the inverse process of step 1~(a). For a phase $\phi$ accumulated in steps 2 and 4 that is a multiple of $\pi$, this results in the population returning to the ground state $\ket{0}$. More generally, the final state is a superposition of $\ket{0}$ and $\ket{1}$, depending on the accumulated phase $\phi$, with the population in $\ket{0}$ as
      \begin{equation}
        \label{eq:P_0}
        P_0(\phi) \equiv \Abs{\Braket{0 | \Psi(T)}}^2 = \cos^2\left(\frac{\phi}{2}\right)
      \end{equation}
      and the population in $\ket{1}$ as
        $P_1(\phi) = 1 - P_0(\phi)$.
    \end{enumerate}

\end{enumerate}

There are possible variations of the above procedure. For example, the interferometer could be designed symmetrically, where, in step 1, the initial state is split into a superposition of $\ket{-N}$ and $\ket{N}$~\cite{McGuirkPRL2000,SaywellPRA2020}. However, this generally requires additional laser fields, as a single pair of counter-propagating fields cannot address both branches of the interferometer.
Furthermore, the subdivision into a splitting between levels $\ket{0}$ and $\ket{1}$, which is then amplified to a superposition between $\ket{0}$ and $\ket{N}$, could be generalized so that the initial splitting is between $\ket{0}$ and some higher level $\ket{N_S}$ with $1 < N_S < N$, and where $\ket{N_S}$ is then still further amplified to $\ket{N}$. Physically, this situation would occur with higher-order Bragg pulses~\cite{ChiowPRL2011}, where the final state of the interferometer would then be a superposition of $\ket{0}$ and $\ket{N_S}$. Here, we consider only first-order Bragg pulses, which change the momentum in steps of $2\hbar k$, following the setup in Reference~\cite{KovachyN2015}.

The interferometer signal contrast is defined as
\begin{equation}
  \label{eq:contrast}
  C \equiv \frac{P_{\max}  - P_{\min}}{P_{\max} + P_{\min}} \,,
\end{equation}
with $P_{\max} \equiv \max_{\phi} P_0(\phi) = P_0(\phi=0)$ and $P_{\min} \equiv \min_{\phi} P_0(\phi) = P_0(\phi=\pi/2)$.  In a non-ideal system, with imperfect splitting and mirror operations, population may be lost from the $\ket{0}$, $\ket{1}$ subspace at final time, and the population in the ground state may deviate from the ideal Equation~\eqref{eq:P_0}, resulting in a loss of contrast.

\subsection{Train of $\pi/2$ and $\pi$ Pulses}

A straightforward approach is to use a train of $\pi$- and $\pi/2$-pulses to manipulate the wave function of the atoms~\cite{ChiowPRL2011,KovachyN2015}. Each pulse in this train uses a fixed laser frequency $\dot\varphi(t) = \omega^{(n_0)}_L$ that is resonant with a specific transition from level $n_0$ to $n_0+1$. The transition is resonant for
\begin{equation}
  \omega^{(n_0)}_L = -\omega_k (2n_0 + 1)\,.
\end{equation}
Plugging this into Equation~\eqref{eq:E_n} produces ($\beta=0$)
\begin{equation}
  \label{eq:E_n_Rabi}
  E_n^{(\Rabi)} = (n^{\prime\,2} - n^\prime) \hbar \omega_k + n_0 (n_0+1) \hbar \omega_k\,,
\end{equation}
with $n^\prime \equiv n-n_0$.  The second term contributes only a global phase and can be dropped.

\begin{figure}[tb]
  \begin{adjustwidth}{-\extralength}{0cm} 
    \includegraphics{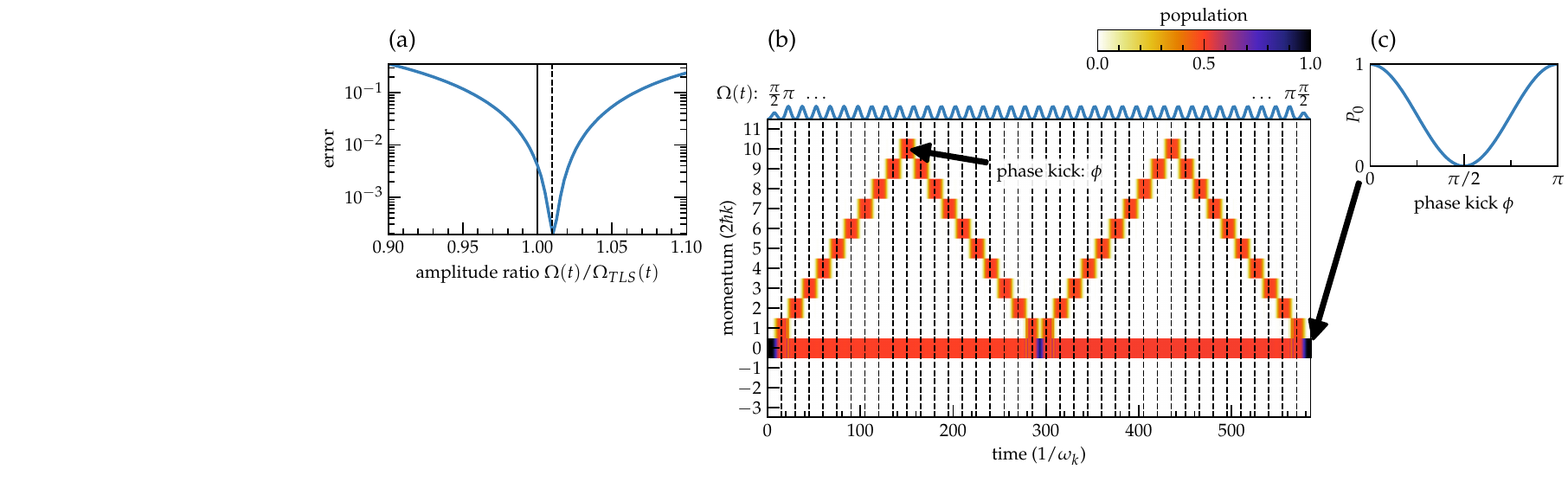}
  \end{adjustwidth}
  \caption{%
    \label{fig:rabi}
    (\textbf{a}) Error $1-\Abs{\Braket{0|\Psi(T)}}^2$ of the population in the ground state after a sequence of $\pi/2$ and $\pi$ pulses for a scaled pulse amplitude. The factor $\Omega(t) / \Omega_{\TLS}(t)$ is the ratio of the pulse amplitude $\Omega(t)$ used in the simulation and the analytic amplitude $\Omega_{\TLS}(t)$ for Rabi cycling in a two-level system with a pulse area of $\pi/2$ or $\pi$, Equation~\eqref{eq:pi_half_condition}. The minimum error is reached for a ratio of $\approx 1.01$, indicated by the dashed vertical line.
    (\textbf{b}) Momentum space dynamics for an interferometric scheme using a sequence of $\pi/2$ and $\pi$ pulses. Each pulse has a Blackman shape, drawn at the top of the panel, with an amplitude adjusted by the correction determined in panel~(a). The time unit $1/\omega_k$ corresponds to roughly $10^{-5}$~s for Rb-87 atoms and a laser wavelength of 780~nm, making the duration of the scheme (excluding the free time evolution) roughly 5.9~ms.
    (\textbf{c}) The final time population in $\ket{0}$ if an instantaneous phase kick is applied to the $\ket{10}$ component of the wave function at maximum separation to account for the free time evolution starting at $t=150\,/\omega_k$ and $t=450\,/\omega_k$ (not shown).
  }
\end{figure}
Steps 1 -- 5 of the full interferometer with a maximum separation of $N=10$ are implemented by a series of $\pi/2$ and $\pi$ pulses, as shown in the top of Figure~\ref{fig:rabi}~(b). Each pulse is a Blackman shape of duration $t_{\Rabi} = 15\,/\omega_k$, which corresponds to roughly 150~\microsec{} for Rb-87 and a laser wavelength of 780~nm. Neglecting all but the two resonant levels for each pulse allows to calculate the pulse amplitude \emph{analytically}; The first pulse is a $\pi/2$ pulse with an amplitude chosen such that
\begin{equation}
  \label{eq:pi_half_condition}
  \int_0^{t_\Rabi}\Omega_{\TLS}(t) dt = \frac{\pi}{2}\,,
\end{equation}
in order to produce a $50/50$ superposition of $\ket{0}$ and $\ket{1}$. The pulse amplitude is $0.125~\omega_k$, corresponding to $1.88 \cdot 2 \pi$\,kHz.

Subsequent pulses are $\pi$-pulses at twice the amplitude of Equation~\eqref{eq:pi_half_condition} that fully transfer population between neighboring levels, respectively swap the population of $\ket{0}$ and $\ket{1}$ at the center of the scheme. The final $\pi/2$ pulse recombines the population.
For simplicity, we do not show the free time evolution in the scheme. The phase that would be accumulated during the free time evolution, when the interferometer is at maximum momentum separation, can be numerically emulated by applying an instantaneous phase kick to the $\ket{10}$-component of the wave function, as indicated in Figure~\ref{fig:rabi}~(b). The shown dynamics are for $\phi = 0$. A kick with a value of $\phi > 0$ only affects the population in the central swap pulse and at final time. The resulting final time population in $\ket{0}$ is shown in Figure~\ref{fig:rabi}~(c) and matches Equation~\eqref{eq:P_0}.

The use of $\pi/2$ and $\pi$ pulses is slightly complicated by the presence of the additional levels.
According to Equation~\eqref{eq:E_n_Rabi}, the transition to the next higher or lower levels is detuned by $2 \omega_k$, which is not completely negligible compared to the amplitude of the $\pi$ pulse, $0.25  \omega_k$. The additional levels induce an effective shift in the two-level system that can be compensated with a modified pulse amplitude. In Figure~\ref{fig:rabi}~(a), we numerically analyze the fidelity of the scheme depending on a scaling factor for the empirical pulse amplitude $\Omega(t)$ relative to the analytical amplitude $\Omega_{\TLS}(t)$ defined in Equation~\eqref{eq:pi_half_condition} for an ideal two-level system. We find that an increase in pulse amplitude by 1\% results in the lowest error. Thus, we include this correction in the scheme shown in Figure~\ref{fig:rabi}~(b) as well as in all future $\pi/2$ and $\pi$ pulse amplitudes.

\subsection{Rapid Adiabatic Passage}

\begin{figure}[tb]
  \begin{adjustwidth}{-\extralength}{0cm} 
    \includegraphics{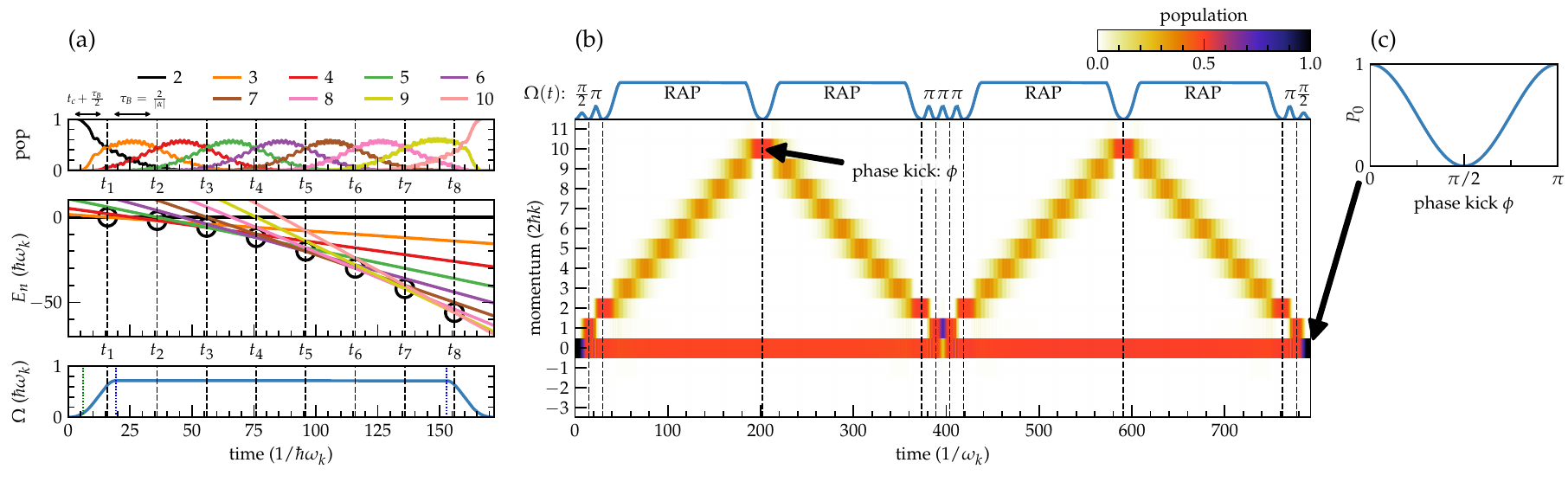}
  \end{adjustwidth}
  \caption{%
    \label{fig:rap}
    (\textbf{a}) Transfer of population from momentum state $\ket{2}$ to $\ket{10}$ using rapid adiabatic passage (RAP) with a constant linear chirp rate $\alpha = 0.1\,\omega_k$ with an offset time of $t_c = 5.93\,/\omega_k$ in Equation~\eqref{eq:E_n_RAP} (green dotted line, bottom). The pulse envelope $\Omega(t)$, shown in the bottom of panel~(a) has a switch-on and switch-off time of $t_r = 19.25\,/\omega_k$ (blue dotted line), using half of a Blackman shape. The center of panel~(a) shows the dynamic energy levels of the Hamiltonian according to Equation~\eqref{eq:E_n_RAP}. Neighboring levels cross at intervals of $\tau_B = 2 / \Abs{\alpha}$, resulting in the transfer of population shown at the top of panel~(a).
    (\textbf{b}) Momentum space dynamics for a full interferometric scheme using an initial $\pi/2$ and $\pi$ pulse to achieve momentum state separation, cf.~Figure~\ref{fig:rabi}. Then, the RAP pulse from panel~(a) first amplifies and then de-amplifies the momentum state separation. The momentum components are swapped with three central $\pi$ pulses. In the second half of the scheme, two additional RAP pulses amplify and de-amplify again. Finally, a $\pi$ and $\pi/2$ pulse perform the recombination. The amplitude of the envelope $\Omega(t)$ in each pulse is drawn to scale at the top of the panel.
    (\textbf{c}) The final time population in $\ket{0}$, if an instantaneous phase kick is applied to the $\ket{10}$ component of the wave function at maximum separation, to account for the free time evolution starting at $t=201.8\,/\omega_k$ and $t=590.6\,/\omega_k$ (not shown).
  }
\end{figure}
An alternative scheme formulated in References~\cite{MalinovskyPRA2003, KovachyPRA2012} is to replace the train of pulses with a single pulse implementing rapid adiabatic passage (RAP) with a linear frequency chirp, respectively, a phase of
\begin{equation}
  \label{eq:phi_rap}
  \varphi_{\RAP}(t) = - \frac{\alpha \omega_k (t-t_c)^2}{2}\,,
\end{equation}
where $\alpha$ is the chirp rate and $t_c$ is a time offset relative to the start of the pulse. Plugging the phase into Equation~\eqref{eq:E_n} with $\beta=0$ results in
\begin{equation}
  \label{eq:E_n_RAP}
  E_n^{\RAP}(t) = (n^2 - n\alpha(t-t_c))\hbar\omega_k\,.
\end{equation}
An example for the dynamic energy levels $E_n^{\RAP}(t)$ and the resulting population dynamics is depicted in Figure~\ref{fig:rap}~(a), with a chirp rate $\alpha = 0.1\,\omega_k$. Neighboring levels cross at time intervals of $\tau_B = 2/|\alpha|$, see the center of panel (a), resulting in a population transfer as shown in the top of panel~(a).

For the given chirp rate, the RAP pulse we employ here is as short as possible, while still transferring momentum with high fidelity. This includes a finite switch on/off from/to zero. The transfer fidelity in this case depends on the specific switch-on time and shape, and an offset $t_c$ in Equation~\eqref{eq:phi_rap} that determines the zero point of the induced energy shift. Effectively, this offset accounts for the fact that the RAP dynamics start only when the field reaches some minimum amplitude. We use a Blackman shape for the switch-on/off and numerically determined the switch-on time $t_r$, the offset $t_c$ and the peak amplitude via an optimization of the transfer fidelity using the Nelder-Mead method. We found $t_c=5.927\,/\omega_k$ for a switch-on/off time of $t_r = 19.252\,/\omega_k$ and a peak amplitude of $0.7\,\omega_k$, corresponding to roughly $11 \cdot 2\pi$~kHz for Rb-87 and a laser wavelength of 780~nm. The resulting envelope $\Omega(t)$ is shown at the bottom of panel~(a) with $t_c$ and $t_r$ indicated by the dotted green and blue lines, respectively.

A noteworthy feature of the RAP population transfer is that it is relatively insensitive to the pulse amplitude, promising some degree of robustness when used as part of an atom interferometric scheme. It also allows to transfer population to arbitrarily high momentum states, for as long as the constant chirp can be maintained. However, compared to the ``Rabi'' scheme consisting only of $\pi/2$ and $\pi$ pulses, it is relatively more challenging to use RAP for the initial splitting, the central swap of $\ket{0}$ and $\ket{1}$, or the final recombination required for the full scheme. Thus, we combine it with a $\pi/2$ pulse to achieve the initial splitting, followed by another $\pi$ pulse to achieve sufficient separation for the RAP scheme to transfer population to maximum separation. In the center of the scheme, three $\pi$ pulses swap the amplitude, and, finally, a $\pi$ pulse and a $\pi/2$ pulse perform the recombination. The full scheme is depicted in Figure~\ref{fig:rap}~(b).

Compared to the sequence of $\pi/2$ and $\pi$ pulses in Figure~\ref{fig:rabi}~(b), the required pulse area for RAP is significantly larger; see the amplitudes shown in the top of both panels. Further compressing the RAP pulses in time would cause increasingly non-adiabatic dynamics and a breakdown of the population transfer, shown in Figure~\ref{fig:rap}~(a). The fast oscillations that can be seen at the top of panel~(a) are already non-adiabatic effects, showing that the chosen parameters approach the time limit for RAP\@. Thus, the RAP scheme in Figure~\ref{fig:rap}~(b) is slightly slower than the comparable Rabi scheme of $\pi/2$ and $\pi$ pulses in Figure\ref{fig:rabi}~(b), with a combined duration $T=792.4\,/\omega_k$ versus $T=585.0\,/\omega_k$. In general, this should have a negligible effect on the overall interferometer, as the free time of flight, measured in seconds~\cite{KovachyN2015}, dominates the time $T$ for splitting, mirroring, and recombination, measured in milliseconds.

\section{Robustness}%
\label{sec:robustness}

The dynamics shown in Figures~\ref{fig:rabi},~\ref{fig:rap} for the analytical Rabi and RAP scheme are for ideal parameters, i.e., zero velocity relative to the rest frame and the ideal pulse amplitude, $\beta = 0$ and $\mu = 1$ in Equations~\eqref{eq:hamiltonian} and~\eqref{eq:E_n}), where $\mu = 1$ now includes the correction of $1.01$, obtained from Figure~\ref{fig:rabi}~(a). We can now analyze the robustness of the different interferometric schemes with respect to variations in atom velocity and pulse amplitude. Since the width of the atomic cloud is small relative to the cross section of the laser pulse, deviations from the ideal amplitude are most likely due to the position of the atomic cloud within the laser field, or due to variations in the overall laser amplitude itself, but are the same for all atoms in the ensemble. On the other hand, the velocity relative to the rest frame varies within the ensemble. It is normal-distributed with a variance that depends on the temperature of the atomic cloud.

\begin{figure*}[tb]
  \includegraphics{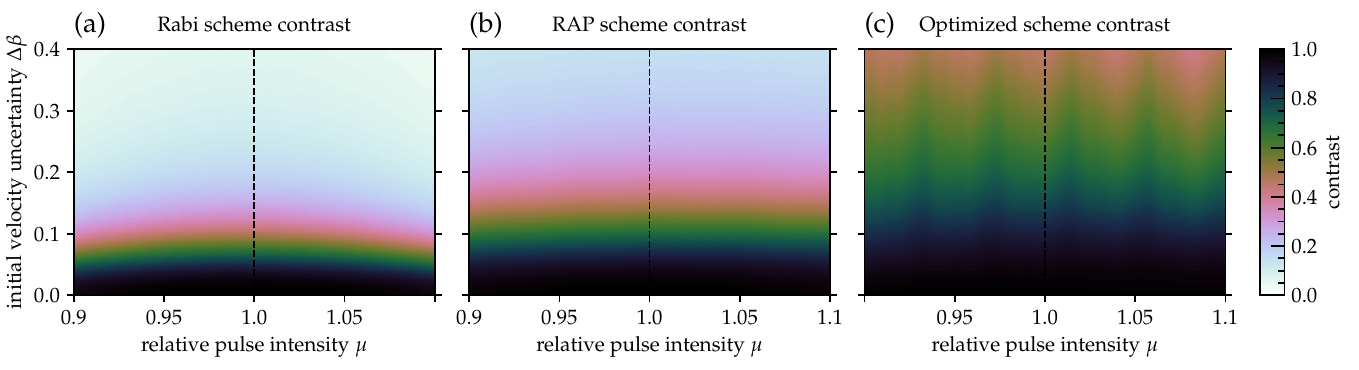}
  \caption{%
    \label{fig:contrast}
    Contrast achieved with both analytical and optimized pulse schemes for a full $20\,\hbar k$ interferometer scheme. In each panel, the expectation value of the signal contrast is shown for a fixed amplitude scaling factor $\mu$ of the ideal pulse amplitude, and assuming a Gaussian distribution with width $\Delta\beta$ for the atom's initial momentum relative to the rest frame in units of $2\hbar k$. The schemes are (\textbf{a}) a train of $\pi/2$ and $\pi$ Rabi pulses, (\textbf{b}) a combination of $\pi/2$ and $\pi$ pulses with rapid adiabatic passage~(RAP), and (\textbf{c}) a scheme using optimized control pulses in combination with rapid adiabatic passage, cf. Figures~\ref{fig:rabi},~\ref{fig:rap},~\ref{fig:oct_dynamics}. The value of the contrast for each point is obtained from the average populations in the ground state, see Equation~\eqref{eq:avg_contrast}.
  }
\end{figure*}
In Figure~\ref{fig:contrast}~(a,~b), we show the expectation value of the contrast for static deviations from the optimal pulse amplitude by $\pm 10\%$ ($\mu \in [0.9, 1.1]$) and for $\beta$ (the momentum relative to the rest frame in units of $2\hbar k$) drawn from a normal distribution with a standard deviation between 0 and $0.4$. For every point in this robustness landscape, we evaluated $N=50\,000$ samples to find
\begin{equation}
  \begin{split}
    \bar{P}_{\max}(\mu, \Delta\beta) &= \frac{1}{N} \sum_{n=1}^N P_0(\phi=0; \mu, \beta_n) \,,\\
    \bar{P}_{\min}(\mu, \Delta\beta) &= \frac{1}{N} \sum_{n=1}^N P_0(\phi=\frac{\pi}{2}; \mu, \beta_n)\,,
  \end{split}
\end{equation}
where $P_0$ is the population in the ground state at final time, $\phi$ is the differential phase accumulated between the two branches of the interferometer, and $\beta_n$ is a value of $\beta$ drawn from the distribution of width $\Delta\beta$. The expectation value of the contrast shown in Figure~\ref{fig:contrast} is then
\begin{equation}
  \label{eq:avg_contrast}
  \bar{C}(\mu, \Delta\beta) = \frac{\bar{P}_{\max}(\mu, \Delta\beta) - \bar{P}_{\min}(\mu, \Delta\beta)}{\bar{P}_{\max}(\mu, \Delta\beta) + \bar{P}_{\min}(\mu, \Delta\beta)}\,,
\end{equation}
cf.~Equation~\eqref{eq:contrast}.

We find that the contrast is relatively robust with respect to deviations from the optimal pulse amplitude, but decays quickly for broader distributions in the atomic velocity. Using the scheme of $\pi/2$ and $\pi$ Rabi pulses, Figure~\ref{fig:rabi}~(b), the contrast crosses the 50\% mark for a standard deviation of $\Delta\beta \approx 0.1$ and effectively approaches zero for $\Delta\beta > 0.2$. Taking advantage of rapid adiabatic passage (RAP) with the scheme shown in Figure~\ref{fig:rap}~(b), we find a measurable improvement in robustness. The sensitivity to deviations in $\mu$ nearly disappears, and the loss of contrast, due to $\Delta\beta > 0$, is reduced by at least a factor of $1.5$. That is, a 50\% loss of contrast occurs at $\Delta\beta \approx 0.15$.  Even at $\Delta\beta=0.4$, the contrast is still $\approx 12\%$.

\begin{figure*}[tb]
  \includegraphics{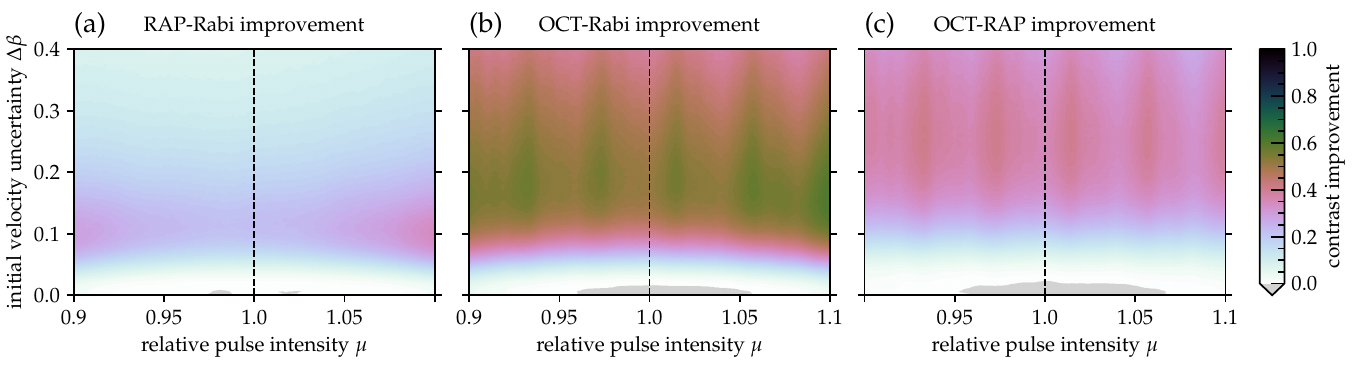}
  \caption{%
    \label{fig:improvement}
    Contrast improvement between different schemes. Panel~(\textbf{a}) shows the difference between Figure~\ref{fig:contrast}~(a,~b), that is, between a scheme using a train of $\pi/2$ and $\pi$ Rabi pulses and a scheme using rapid adiabatic passage (RAP). Panels~(\textbf{b}, \textbf{c}) show the difference between Figure~\ref{fig:contrast}~(a,~c), respectively Figure~\ref{fig:contrast}~(b,~c); that is, between a scheme using pulses derived from optimal control theory (OCT) and the two analytic schemes (Rabi, RAP). The light gray points mark a (negligible) loss of contrast, $\Abs{\Delta C} < 0.04$ in panel~(a) and $\Abs{\Delta C} < 0.01$ in panels~(b, c).
  }
\end{figure*}
The change in contrast between the Rabi and RAP schemes is quantified in Figure~\ref{fig:improvement}~(a).  The RAP scheme improves on the Rabi scheme by an increase in contrast of up to 0.35. This maximum improvement is reached for deviations of the pulse amplitude near 10\% and a standard deviation of $0.1 \cdot 2 \hbar k$ in the initial momentum of the atoms in the ensemble. The light gray areas in the plot mark a loss of contrast for some points near $\Delta\beta=0$. These losses are comparatively negligible at $\Abs{\Delta C} < 0.04$.

\section{Optimal Control for Robust Pulse Schemes}%
\label{sec:optrobust}

To further increase the robustness, we now consider the use of optimal control theory (OCT). We optimize the specific steps of the interferometric scheme in Section~\ref{sec:analytical} separately: the initial splitting into a superposition of $\ket{0}$ and $\ket{1}$, step 1~(a), the amplification from $\ket{1}$ to $\ket{10}$, steps 1~(b) and 3~(c), the de-amplification, steps 2~(a) and 5~(a), and the swap of amplitudes $\ket{0}$ and $\ket{1}$, step 3~(b).

It can be shown~\cite{GoerzSPIEO2021} that any relative phase introduced by the amplification and de-amplification cancels out. Thus, these steps can be implemented with an optimization functional that only considers populations, e.g., for the amplification step,
\begin{equation}
  \label{eq:J_pop}
  J_{\pop}(\ket{\Psi(T)}) = 1 - \frac{1}{2} \Norm{\vec{P}(\ket{\Psi(T)}) - \vec{P}^{\target}}^2\,,
\end{equation}
where the components of the vectors $\vec{P}$ and $\vec{P}^{\target}$ are the populations in the different momentum levels for the propagated state and the target state, respectively. A relative phase introduced by the initial splitting has to be compensated for in the recombination step. This is automatic if we perform the initial splitting between levels $\ket{0}$ and $\ket{1}$ by optimizing for an effective $\pi/2$ pulse,
\begin{equation}
  \ket{0} \rightarrow \frac{1}{\sqrt{2}}\left(\ket{0} + i \ket{1}\right)\,, \qquad
  \ket{1} \rightarrow \frac{1}{\sqrt{2}}\left(i \ket{0} + \ket{1}\right)
\end{equation}
up to a global phase, i.e., using a square-modulus overlap functional~\cite{PalaoPRA2003}.
This ensures that the same optimized pulse targeting step 1~(a) in Section~\ref{sec:analytical} can also be used for the final recombination, step 5~(b).

For the amplification and de-amplification, we optimized starting from a RAP pulse that transfers $\ket{1} \rightarrow \ket{10}$, respectively $\ket{10} \rightarrow \ket{1}$, cf.~Figure~\ref{fig:rap}~(a). The optimization modifies the envelope $\Omega(t)$ in order to minimize the population functional in Equation~\eqref{eq:J_pop}. In principle, the chirp rate $\alpha$ in Equation~\eqref{eq:phi_rap} could also be made time-dependent. Instead, we left the chirp rate constant and allowed $\Omega(t)$ to be complex-valued.

\begin{figure}[tb]
  \includegraphics{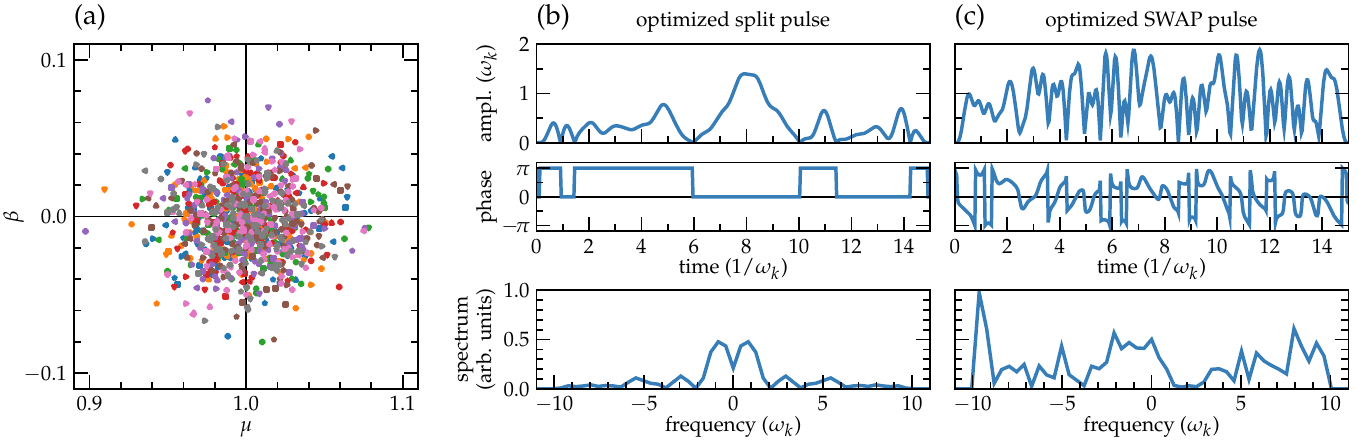}
  \caption{%
    \label{fig:oct_pulses}
    (\textbf{a}) Ensemble points used for the optimization. The sampling points were chosen from a normal distribution around $\mu=1$ and $\beta=0$ with a width of $\Delta\mu=\Delta\beta=0.025$, divided into 64 batches with 16 points per batch. The different batches are distinguished by the combination of color and marker shape. (\textbf{b}) Optimized pulse amplitude, phase, and spectrum for the initial splitting pulse $\ket{0} \rightarrow (\ket{0} + i \ket{1}) / \sqrt{2}$. (\textbf{c}) Optimized pulse amplitude, phase, and spectrum for the central swap pulse between $\ket{0}$ and $\ket{1}$. For Rb-87 and a laser wavelength of 780~nm, the two-photon recoil frequency is $\omega_k = 2 \pi \cdot 15.1$~kHz. The unit of time $1/\omega_k$ corresponds to roughly $10^{-5}$~s. Thus, the duration of the shown pulses is on the order of 150~\microsec.
  }
\end{figure}
To make the optimized pulses robust with respect to deviations in the pulse amplitude and variations in the initial velocity, we employed an ensemble optimization~\cite{LiITAC2009,ChenPRA2014,GoerzPRA2014}. That is, we considered multiple copies of the Hamiltonian, Equation~\eqref{eq:hamiltonian}, each with different parameters $\mu$ and $\beta$. We then optimized over the average of the ensemble. In order to cover a large area of the parameter landscape, we considered an ensemble of 1024 points, split into 64 batches of 16 points each. Individual points were drawn randomly from a normal distribution around $\beta=0$ and $\mu = 1$ with $\Delta\beta = \Delta\mu = 0.025$. For each batch of points, we performed an ensemble optimization with Krotov's method~\cite{TannorBookChapter1992,SomloiCP1993,ReichJCP2012,GoerzSPP2019, GoerzSPIEO2021} for 1000 iterations before moving to the next batch. The procedure continued to loop around the batches until convergence was reached, that is, there was no significant improvement in the fidelity reachable within 1000 iterations, compared to the previous batch.

The sampling points $\mu$ and $\beta$ for the different batches are shown in Figure~\ref{fig:oct_pulses}~(a). We found the chosen width of the sampling point distribution $\Delta\beta = \Delta\mu = 0.025$ to be the maximum width for which an average fidelity on the order of $10^{-3}$ is achievable for the individual components of the interferometer.

As we wanted to explore the limits of robustness achievable via optimal control, we did not restrict the optimized control fields to amplitudes or spectral widths that are easily obtained with current experimental setups. At the same time, we wanted to avoid entirely unrealistic parameter regimes. Thus, we placed a bound on the pulse amplitude at $\Omega_{\max} = 1.5\,\omega_k$, roughly a factor of six higher than the typical amplitude required for a $\pi$ Rabi pulse, and roughly twice the amplitude of the RAP pulses used for the momentum transfer. Similarly, the spectral width was limited to $10\,\omega_k$. Both the amplitude and the spectral width are well within an order of magnitude of the current capabilities of the atomic fountain experimental setup.

The resulting pulses for the initial splitting (an effective $\pi/2$ pulse), and for the swap are shown in Figure~\ref{fig:oct_pulses}~(b,~c). The effective $\pi/2$ pulse is a relatively simple pulse shape. In particular, it does not require a time-dependent phase, i.e., the imaginary part of the control field $\Omega(t)$. The swap is more difficult to realize, and saturates the amplitude and spectral limits placed on the control that allow to reach the almost perfect gate fidelity.

For the optimization of the RAP pulse that amplifies and de-amplifies the momentum state separation, we found that the optimization only added negligible corrections to the pulse shape. This is a testament to the inherent robustness of rapid adiabatic passage. In fact, when combining the optimized components of the interferometer into a full scheme, and analyzing the resulting robustness, we observed no clear advantage in using the RAP pulses with an optimized amplitude. Thus, the full ``optimized'' pulse scheme uses the pulse shown in Figure~\ref{fig:oct_pulses}~(b) as the initial and final component, and the pulse in Figure~\ref{fig:oct_pulses}~(c) as the center swap, but analytical RAP pulses otherwise.

\begin{figure}[tb]
  \begin{center}
    \includegraphics{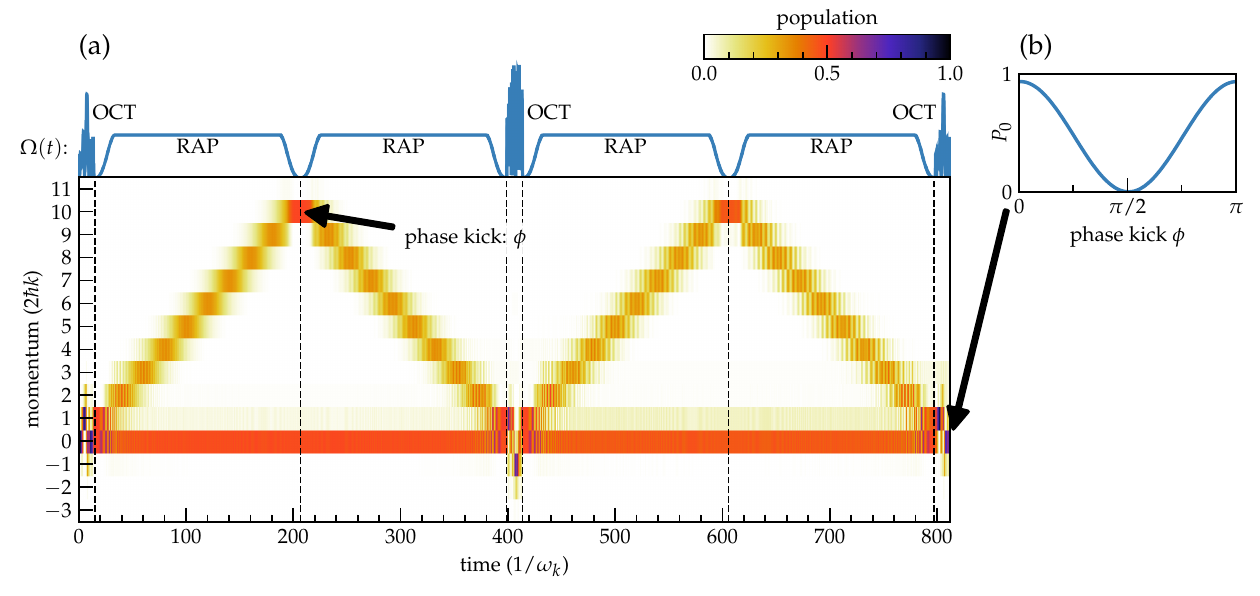}
  \end{center}
  \caption{%
    \label{fig:oct_dynamics}
    (\textbf{a}) Momentum space dynamics for an interferometric scheme using optimized pulses (OCT) in combination with rapid adiabatic passage (RAP). The optimized pulses are those shown in Figure~\ref{fig:oct_pulses}~(b, c) and implement the initial splitting, the center swap, and the final recombination between levels $\ket{0}$ and $\ket{1}$. These are combined with RAP pulses similar to the one shown in Figure~\ref{fig:rap}~(a) transferring population between $\ket{1}$ and $\ket{10}$. All pulse amplitudes are shown to scale at the top of the panel.
    (\textbf{b}) The final time population in $\ket{0}$ if an instantaneous phase kick is applied to the $\ket{10}$ component of the wave function at maximum separation to account for the free time evolution starting at $t=206.8\,/\omega_k$ and $t=605.6\,/\omega_k$ (not shown).
    The maximum population at $\phi=0$ or $\phi=\pi$ is 0.934 and the minimum population at $\phi = \frac{\pi}{2}$ is 0.001.
  }
\end{figure}
The full optimized scheme and the resulting dynamics in the ideal case ($\mu=1$, $\beta=0$) are shown in Figure~\ref{fig:oct_dynamics}. We can see that there are visible deviations from the simple analytic schemes in Figures~\ref{fig:rabi},~\ref{fig:rap}. In particular, the optimized splitting and swap pulses populate outside the two-level subspace $\ket{0}$, $\ket{1}$, at intermediary times. The final time population in $\ket{0}$ differs measurably from the analytical schemes, reaching 0.93 in Figure~\ref{fig:oct_dynamics}~(c). However, since the population for a differential phase of $\phi = \pi/2$ is still 0.001, this does not affect \emph{contrast}, which is still $\approx 1$, according to Equation~\eqref{eq:contrast}.

The contrast of the full scheme for values of $\mu \neq 1$ and for an initial momentum drawn from a normal distribution with $\Delta\beta > 0$ is shown in Figure~\ref{fig:contrast}~(c). We observe a considerable improvement. The limit for 50\% contrast is pushed well beyond $\Delta\beta = 0.3$. Even for $\Delta\beta=0.4$, the minimum contrast is still 41\% or higher. Remarkably, the enhancement in robustness extends far beyond the value of $\Delta\beta = 0.025$ that was used in the ensemble optimization. The improvement relative to the fully analytical Rabi and RAP schemes is shown in Figure~\ref{fig:improvement}~(b,~c). Within the explored parameter regime, the maximum absolute improvements in contrast are 0.61 and 0.41, respectively. The losses marked in light gray are $\Abs{\Delta C} < 0.01$ in both cases.

\section{Conclusion and Outlook}%
\label{sec:conclusion}

We have numerically analyzed the expected robustness of several complete schemes for an atomic fountain interferometer reaching a momentum state separation of $20\,\hbar k$. For purely analytic schemes, we found that robustness can be increased considerably by using rapid adiabatic passage to transfer population after the initial separation. This comes at the cost of an increase in pulse area, reaching about the limit of current experimental capabilities. However, the use of rapid adiabatic passage fundamentally enables the achievement of arbitrarily high momentum state separation and preserves very high robustness, as long as the linear chirp of the laser frequency can be maintained. In fact, we did not find that the robustness of the RAP pulses can be substantially improved by optimal control. In contrast, optimal control theory can significantly improve the initial splitting and swapping of amplitudes in the middle of the interferometric scheme. Again, this comes at the cost of an increase in pulse area. The optimized pulses presented here are on the border of current experimental capabilities in terms of pulse amplitude and spectral width, but well within an order of magnitude. Thus, we expect these pulses to be realizable in the future.

Combining optimized pulses with rapid adiabatic passage into a full scheme results in a very robust scheme, if the uncertainty of the initial momentum of the atoms in the interferometer is $>0.1 \cdot 2 \hbar k$. Even for relatively large uncertainties of $0.4 \cdot 2 \hbar k$ or higher, a contrast of 40\% is maintained. Due to the combination with analytic RAP pulses, we expect this contrast to be maintained even for much higher momentum state separation.

The optimized scheme identified here opens several avenues for future exploration. For implementation in the laboratory, the pulse area and spectral width of the optimized pulses would have to be reduced by at least a factor of three. As the goal here was to identify maximally robust pulses without stringent constraints, the limits of robustness within currently achievable constraints of a specific experimental setup have not been fully probed. As RAP was identified here as a core component of a robust scheme with large momentum state separation, it would be worthwhile to consider the combination of RAP with other two-level control schemes, such as those derived from nuclear magnetic resonance~\cite{DunningPRA2014}.

We have assumed here that the deviations in the pulse amplitude are homogeneous, i.e., the atomic cloud is small relative to the cross section of the laser. Further, we have assumed that there are no time-dependent fluctuations in the laser for the duration of the pulse scheme, on the order of milliseconds. Strategies for mitigating time-dependent noise will be considered in future work. Spatial distortions in the laser profile could be taken into account by extending the model beyond plane waves~\cite{FitzekSR2020}.

More generally, the sensitivity of the interferometer could be enhanced by exploiting correlation between the atoms, e.g., spin squeezing~\cite{PezzeRMP2018, BrifPRXQ2020,GreveN2022,MaliaN2022}. We have applied optimal control to the creation of such squeezed states~\cite{CarrascoPRA2022}. Going forward, we would like to explore the use of optimal control to further enhance the robustness of correlated atoms in atom interferometric schemes, including alternative realizations, such as lattice guided~\cite{KovachyPRA2010} and tractor atom interferometers~\cite{RaithelQST2022}.

\section*{Acknowledgments}%

The authors thank Remy Notermans, Chris Overstreet, Peter Asenbaum, Tim Kovachy, and Sebastián Carrasco for fruitful discussions. This work was partially supported by the ECI-DIRA program at the DEVCOM Army Research Laboratory. MHG acknowledges support by the DEVCOM Army Research Laboratory under Cooperative Agreement Number W911NF-16-2-0147. MAK was supported by the U.S. Department of Energy, Office of Science, National Quantum Information Science Research Centers, Superconducting Quantum Materials and Systems Center (SQMS) under contract number DE-AC02-07CH11359.

\reftitle{References}

\bibliography{refs}

\end{document}